\documentclass[oneside,12pt]{article}
\usepackage{graphics,color,epsfig}
\usepackage{natbib}

\newcommand{\kms}{\,{\rm km/s}}
\newcommand{\micron}{\mu{\rm m}}
\newcommand{\mysection}[1]{\medskip\noindent{\large\bf #1}\par\smallskip}

\topmargin = -\topskip
\advance\topmargin by -\headsep
\begin{document}

\pagestyle{empty}

\large
\centerline{\bf SEGUE-2 and APOGEE: Revealing the History of the Milky Way}
\normalsize

\bigskip
\bigskip
\centerline
{White Paper for the Astro2010 GAN Science Frontier Panel}



\begin{itemize}

\item Constance Rockosi, U.C. Santa Cruz, crockosi@ucolick.org \\
SEGUE-2 Principal Investigator

\item Timothy C. Beers, Michigan State University, beers@pa.msu.edu \\
SEGUE-2 Survey Scientist  

\item Steven Majewski, University of Virginia, srm4n@mail.astro.virginia.edu\\
APOGEE Principal Investigator

\item Ricardo Schiavon, Gemini Observatory, rschiavon@gemini.edu\\
APOGEE Survey Scientist

\item Daniel Eisenstein, University of Arizona, deisenstein@as.arizona.edu \\
SDSS-III Director

\end{itemize}

\bigskip
\centerline{\bf ABSTRACT}

The history of the Milky Way is encoded in the spatial distributions,
kinematics, and chemical enrichment patterns of its resolved stellar
populations. SEGUE-2 and APOGEE, two of the four surveys that comprise SDSS-III
(the Sloan Digital Sky Survey III), will map these distributions and enrichment
patterns at optical and infrared wavelengths, respectively. Using the existing
SDSS spectrographs, SEGUE-2 will obtain spectra of 140,000 stars in selected
high-latitude fields to a magnitude limit $r \approx 19.5$, more than doubling
the sample of distant halo stars observed in the SDSS-II survey SEGUE (the Sloan
Extension for Galactic Understanding and Exploration). With spectral resolution
$R \approx 2000$ and typical S/N per pixel of $20-25$, SEGUE and SEGUE-2 measure
radial velocities with typical precision of $5-10\kms$ and metallicities
([Fe/H]) with a typical external error of 0.25 dex. APOGEE (the Apache Point
Observatory Galactic Evolution Experiment) will use a new, 300-fiber $H$-band
spectrograph ($\lambda = 1.5-1.7\micron$) to obtain high-resolution ($R \approx
24,000$), high signal-to-noise ratio (S/N$\approx 100$ per pixel) spectra of
100,000 red giant stars to a magnitude limit $H \approx 12.5$. Infrared
spectroscopy penetrates the dust that obscures the inner Galaxy from our view,
allowing APOGEE to carry out the first large, homogeneous spectroscopic survey
of {\it all} Galactic stellar populations. APOGEE spectra will allow radial
velocity measurements with $< 0.5$ $\kms$ precision and abundance determinations
(with $\sim 0.1$ dex precision) of 15 chemical elements for each program
star, which can be used to reconstruct the history of star formation that
produced these elements. Together, SEGUE-2 and APOGEE will provide a picture of
the Milky Way that is unprecedented in scope, richness, and detail. The combined
data set will play a central role in ``near-field cosmology'' tests of galaxy
formation physics and the small scale distribution of dark matter.


\clearpage

\pagestyle{plain}
\setcounter{page}{1}

\mysection{1. Introduction}

Observations over the past decade have demonstrated that the Milky Way is
spatially and kinematically complex, and includes numerous lumps and streams in
the stellar halo (which itself can be resolved into two distinct components, an
inner and outer halo), substructures in the stellar disk systems, and dwarf
satellites with an enormous range of luminosities. Cold dark matter (CDM)
cosmological models predict complex assembly histories for galaxies like the
Milky Way, with many or most spheroid stars formed in smaller systems that have
since been disrupted, and with the stellar disk frequently perturbed by minor
mergers and dynamical interactions with halo substructures. While some of these
predictions are supported by the observations, the physics of disk galaxy
formation in CDM remains poorly understood. Large stellar spectroscopic surveys
can map the spatial and kinematical structure of the Galaxy, and probe its star
formation history by measuring the chemical enrichment patterns in today's
stellar populations. Abundances in the most chemically primitive spheroid stars
allow ``archaeological'' investigations of the earliest epochs of cosmic star
formation, whose nucleosynthetic products enriched these stars. In short,
massive spectroscopic surveys of the Milky Way do more than simply map our
Galactic home -- they open a new frontier in the study of galaxy formation,
dark matter, stellar nucleosynthesis, and star formation in the early Universe.

Many recent advances in Galactic structure have come from the Sloan Digital Sky
Survey (SDSS; York et al.\ 2000), especially the SEGUE (Sloan Extension for
Galactic Understanding and Exploration) survey of SDSS-II. Using the existing
SDSS spectrographs, SEGUE-2 will obtain medium-resolution (R = 2000) spectra of
an additional 140,000 stars in selected high-latitude fields to a magnitude
limit $r \approx 19.5$, more than doubling the sample of distant halo stars
observed in the SEGUE.  These two surveys will provide maps of the kinematics and
metallicities of stellar populations over a large volume of the Galaxy, from the
inner halo to large Galactocentric distances where late-time accretion events
are expected to dominate the outer-halo population.  

The Apache Point Observatory Galactic Evolution Experiment (APOGEE) will use a
new, 300-fiber $H$-band spectrograph ($\lambda =
1.5-1.7\micron$) to carry out a high-resolution ($R\approx 24,000$)
spectroscopic survey of $10^5$ red giant stars to a magnitude limit $H \approx
12.5$, penetrating interstellar dust obscuration to provide the first large
scale, uniform, high precision spectroscopic study of {\it all} Galactic stellar
populations (bulge, disk, bar, halo). APOGEE spectra will allow radial velocity
measurements with $< 0.5$ $\kms$ precision and abundance determinations (with $\sim
0.1$ dex precision) of 15 chemical elements for each program star, which
can be used to reconstruct the history of star formation that produced these
elements, and reveal subtle kinematic disturbances or pinpoint kinematic
substructures. Together, SEGUE-2 and APOGEE will provide a picture of the Milky
Way that is unprecedented in scope, richness, and detail. The combined data set
will play a central role in ``near-field cosmology'' tests of galaxy formation
physics and the small scale distribution of dark matter.

SEGUE-2 and APOGEE are part of SDSS-III, a six-year program (2008-2014) that
will use existing and new instruments on the 2.5-m Sloan Foundation Telescope to
carry out four spectroscopic surveys on three scientific themes: dark energy and
cosmological parameters; the structure, dynamics, and chemical evolution of the
Milky Way; and the architecture of planetary systems.\footnote{Because the
different SDSS-III surveys are relevant to different Astro2010 survey panels, we
are providing three separate White Papers, though the general material on the
SDSS is repeated. A detailed description of SDSS-III is available at {\tt
http://www.sdss3.org/collaboration/description.pdf}.} All data from SDSS-I
(2000-2005) and SDSS-II (2005-2008), fully calibrated and accessible through
efficient data bases, have been made public in a series of roughly annual data
releases, and SDSS-III will continue that tradition. SDSS data have supported
fundamental work across an extraordinary range of astronomical disciplines,
including the large-scale structure of the universe, the evolution and
clustering of quasars, gravitational lensing, the properties of galaxies, the
members of the Local Group, the structure and stellar populations of the Milky
Way, stellar astrophysics, sub-stellar objects, and small bodies in the solar
system. A summary of some of the major scientific contributions of the SDSS to
date appears in the Appendix.

By the time
of the Astro2010 report, we hope that SDSS-III fundraising will be complete.
We are providing SDSS-III White Papers to the Astro2010 committee and panels
mainly as information about what we expect to be a major activity for the first
half of the next decade, and about datasets that will shape the environment for
other activities. We also wish to emphasize the value of supporting projects of
this scale, which may involve public/private partnerships and international
collaborations like the SDSS, and thus the importance of maintaining funds and
mechanisms to support the most meritorious proposals that may come forward in
the next decade.

\mysection{2. Description of SEGUE-2 and APOGEE}

SEGUE-2 is a continuation of SEGUE, similar in overall strategy but with targeting
designed to focus more on the outer-halo population. The 240,000 targets
in SEGUE emphasized stars in the relatively nearby regions of the Galaxy, with
the majority of the fibers placed on stars expected to be within 10 kpc,
including a substantial number of thick-disk and inner-halo stars  (Yanny et al.
2009). SEGUE-2 will survey the Galaxy's stellar halo \textit{in situ} at
distances from 10 kpc to beyond 60 kpc. The goal is to use the combination of
SEGUE and SEGUE-2 to map the kinematics and stellar populations over a large
volume of the Galaxy, from the inner halo to the outer halo. Ultimately, this
information will be used to constrain existing models for the formation of the
stellar halo(s), and to inform the next generation of high-resolution
simulations of galaxy formation.   

The outer halo has already been shown to be of particular interest, as analysis
of SDSS/SEGUE spectroscopy revealed that ``the halo'' of the Galaxy can be
resolved into at least two distinct components with markedly different
distributions of (1) metallicity (the outer-halo metallicity distribution
function, MDF, peaking at [Fe/H] = $-2.2$, a factor of 3 lower than that of the inner
halo), (2) kinematics (the outer-halo exhibiting a net retrograde rotation on
the order of $-70$ to $-80$ km/s, compared to a zero or slightly prograde net
rotation of the inner halo), and (3) density (the outer halo exhibiting rounder
spatial density contours, compared to a moderately flattened inner halo). See
Carollo et al. (2007) for additional details. The desire to better constrain the
nature of the inner- and outer-halo populations, and to examine the critical
transition region from the populations, which occurs between 15 and 20 kpc from
the Galactic center, strongly influences the targeting philosophy of SEGUE-2.    

The roughly 140,000 stars targeted by SEGUE-2 are selected from SDSS imaging,
isolated by a variety of color cuts (and in some cases, mild proper
motion cuts) to populate the primary selection categories. The categories
include blue horizontal-branch stars, K- and M-type giants, and an unbiased
(with respect to metallicity and kinematics) sample of F-turnoff stars. SEGUE-2
also includes categories for very metal-poor stars, extreme and ultra-cool
subdwarfs, and other rare but interesting subsamples, such as hyper-velocity
star candidates. In selecting distant and rare objects, SEGUE-2 benefits from the
experience of SEGUE in developing optimized and efficient selection techniques,
setting magnitude limits to balance survey depth, sky coverage, and data quality,
and in having a stable stellar parameters pipeline (the SDSS/SEGUE Stellar
Parameters Pipeline, SSPP; see Lee et al. 2008a,b and Allende Prieto et al.
2008), with well-understood errors as a function of color and signal to noise.

SEGUE-2 employs the existing SDSS spectrographs (which provide wavelength
coverage 3800 - 9200 {\AA}, at R = 2000), with integration times set to achieve
an average S/N per pixel of 20/1. With these data, the SSPP achieves typical
external errors of $5-10\kms$, 140~K, 0.29 dex, 0.24 dex for radial velocity,
$T_{\rm eff}$, log$g$, and [Fe/H], respectively. The combined SEGUE+SEGUE-2 data
set is expected to approach 400,000 stars, with more than 100,000 of these
targets sampling the outer-halo population (doubling the SEGUE sample of such
distant stars). When combined with stellar observations obtained as part of the
SDSS-I/SDSS-II Legacy programs, the final sample should be on the order of
500,000 stars.   

The anticipated science outcomes of the combined SEGUE/SEGUE-2 samples include:
\begin{itemize}

\item Greatly enlarged samples to enable the detection and analysis of
individual stellar streams
in the inner and outer halo (e.g., Harrigan et al. 2009; Klement et al. 2009;
Willet et al. 2009), as well as the ``global'' level of kinematic substructure
present throughout the Galaxy (e.g., Schlaufman et al., in preparation). The
velocity and metallicity accuracies achieved by SEGUE/SEGUE-2 are sufficient for
first-pass identification, which can be followed up (for nearby streams) with
APOGEE and (for more distant streams) with other surveys. 

\item Expanded samples of distant blue horizontal-branch stars and M-giants, which
enable refined estimates of the mass (and mass profile) of the Galaxy (e.g., Xue
et al. 2008), as well as {\it in-situ} studies of the change of the halo MDF
with distance.   

\item Assembly of a sample of over 20,000 stars with [Fe/H] $<
-2.0$, enabling a definitive study of the shape of the low-metallicity tail of the
inner- and outer-halo MDFs (e.g., Beers et al. 2009). The lowest metallicity stars
identified are, and will continue to be, studied with high-resolution
spectroscopy using existing 8-10m telescopes, and in the future, with Extremely
Large Telescopes. Such studies are invaluable for revealing the elemental
abundance patterns that were produced by the first generations of stars.

\item Determination of the velocity ellipsoids and
relative normalizations of the thick-disk, metal-weak thick-disk, inner- and
outer-halo populations. This goal, already underway with existing data from
SDSS/SEGUE (Carollo et al., in preparation), provides the required constraints
for new generation, cosmologically tuned models of galaxy formation and
evolution (e.g., Tumlinson 2006).

\item The identification and kinematic study of many thousands of extreme sdM stars in the local
volume, enabling explorations of the low-mass stars of the halo populations
(e.g., Lepine et al. 2009). 

\end{itemize}

APOGEE will produce the first systematic survey of the {\it 3-D distribution
functions} of the abundances of 15 chemical elements that are key for the
understanding of the star formation and chemical evolution of the Galaxy. This
will be achieved by use of a new 300-fiber cryogenic high-resolution
spectrograph that will provide access to regions of high extinction in the
Galactic inner disk and bulge. Accurate abundances (0.1 dex) will be obtained
for elements such as oxygen, carbon, and nitrogen, which are the most abundant
metals and preferred chemical evolution tracers, as well as other metals with
particular sensitivity to the star formation history and the initial mass
function, such as iron-peak, odd-Z, and $\alpha$-elements. APOGEE will also
obtain accurate (0.5 km/s) kinematical data useful for constraining dynamical
models for the disk, the bulge, the bar and the halo, and for discriminating
substructures in these components, if/where they exist. The enormous APOGEE data
set will make possible the determination of metallicity and abundance pattern
distribution functions for many dozens of Galactic zones ($R$, $\theta$, $Z$) at
the level of detail currently available only for the solar neighborhood.

Some of the prime science goals of APOGEE include:

\begin{itemize} 

\item To constrain models of hierarchical formation of the Galaxy
by searching for residues of merging processes in the form
of velocity substructure and/or through chemical fingerprinting, thus establishing
the relative contributions of {\it in situ} star formation and accretion of
previously formed stellar populations. APOGEE will increase, by
orders of magnitude, the samples of bulge stars with accurately determined
abundance patterns and kinematics, which will place definitive constraints on its
accretion and star formation history.

\item To identify low-latitude halo sub-structure like the Monoceros
stream, and to determine velocity dispersions of the currently known
tidal streams, in order to constrain the mass of the disrupted
parent satellites and to place limits on dynamical heating of streams
by lumps in the dark matter halo.

\item To constrain the mass profile of the Galaxy, by combining radial
velocities with spectroscopic parallaxes to map the large scale dynamics of
the bulge, bar, and disk, thereby probing the global distribution of light
and dark matter via the first comprehensive determination of the rotation
curve to the outermost reaches of the disk.

\item To infer properties of Population III stars (thought to reside
or to have resided in the Galactic bulge) by detecting them directly
if they survive to the present day, or by measuring their nucleosynthetic
products in the most metal-poor stars surviving to the present day.

\item To carry out a legacy survey of low-latitude star clusters, which will
provide strong constraints on the history of star formation and chemical
enrichment of the inner Galaxy, via the combination of detailed abundance
patterns with CMD-based ages.

\item To constrain the high-mass end of the initial mass function in early
stellar populations at different Galactic zones from abundance ratios
sensitive to yields from stars of different masses.

\item To characterize the dynamics and chemistry of the Galactic bar from
the first major survey of its stars.  Effects of the bar on the dynamics of
the disk and nearby halo will also be studied.

\end{itemize}

\mysection{3. Conclusions}

SEGUE/SEGUE-2 (along with RAVE) are the first of numerous observational campaigns
over the coming decade that will revolutionize our understanding of the history
of the formation and evolution of large galaxies such as the Milky Way. Future
surveys include Galactic studies to be conducted by LAMOST in China, HERMES at
AAO, WFMOS at Gemini/Subaru, ESA's GAIA satellite, Pan-STARRS in Hawaii,
SkyMapper in Australia, and the LSST. The ground-breaking work of the SDSS survey
efforts will inform and greatly improve the scientific return from all of these
future missions.   

The uniqueness of APOGEE emanates from the fact that it is a {\it
high-resolution, near-infrared} survey of a {\it huge ($10^5$)} sample of
Galactic stars. High-resolution spectroscopy will make possible the
determination of very accurate elemental abundances via the application of
classical abundance analysis methods to a large stellar sample in an automated
fashion. Moreover, the significantly reduced extinction in the H-band compared
to optical wavelengths ($A_H = A_V / 6$) will allow APOGEE to reach uncharted
territory, well beyond distances accessible to any of the other major surveys
coming to fruition in the next decade. Finally, in keeping with SDSS tradition,
the immense homogeneous database produced by APOGEE is poised to spark numerous
follow-up observational pursuits aimed at further exploring the many
avenues that it will open for investigation.

In round numbers, SDSS-III is a \$40 million project, and the funding is largely
in hand thanks to generous support from the Alfred P. Sloan Foundation, the
National Science Foundation, the Department of Energy, and the Participating
Institutions (including international institutions and participation groups
supported, in some cases, by their own national funding agencies). 
The SDSS has demonstrated the great value of homogeneous surveys that provide
large, well-defined, well-calibrated data sets to the astronomical community. In
many cases, such surveys are made possible by novel instrumentation, and they
often require multi-institutional teams to carry them out. The case for
supporting ambitious surveys in the next decade is best made by considering the
contributions of the SDSS to the astronomical breakthroughs of the {\it current}
decade, as summarized in the Appendix below.

\clearpage
\large\noindent
{\bf Appendix: The SDSS Legacy}
\normalsize

The SDSS (York et al.\ 2000) is one of the most ambitious and influential
surveys in the history of astronomy.
SDSS-II itself comprised three surveys: the Sloan Legacy Survey completed
the goals of SDSS-I, with imaging of 8,400 square degrees and spectra of
930,000 galaxies and 120,000 quasars; the Sloan Extension for Galactic 
Understanding and Exploration (SEGUE) obtained 3500 square degrees of 
additional imaging and spectra of 240,000 stars; and the Sloan Supernova Survey
carried out repeat imaging ($\sim 80$ epochs) of a 300-square degree area,
discovering nearly 500 spectroscopically confirmed Type Ia supernovae for
measuring the cosmic expansion history at redshifts $0.1 < z < 0.4$.
Based on an analysis of highly cited papers, Madrid \& Machetto
(2006, 2009) rated the SDSS as the highest impact astronomical
observatory in 2003, 2004, and 2006 (the latest year analyzed so far).
The final data release from SDSS-II was made public in October, 2008,
so most analyses of the final data sets are yet to come.

The list of extraordinary scientific contributions of the SDSS
includes, in approximately chronological order:
\begin{itemize}
\item{} 
{\it The discovery of the most distant quasars,}
tracing the growth of the first supermassive black holes and
probing the epoch of reionization.
\item{} 
{\it The discovery of large populations of L and T dwarfs,}
providing, together with 2MASS, the main data samples for systematic
study of sub-stellar objects.
\item{} 
{\it Mapping extended mass distributions around galaxies with weak
gravitational lensing,} demonstrating that dark matter halos extend to 
several hundred kpc and join smoothly onto the larger scale dark matter 
distribution.
\item{} 
{\it Systematic characterization of the galaxy population,}
transforming the study of 
galaxy properties and the correlations among them
into a precise statistical science, yielding powerful insights
into the physical processes that govern galaxy formation.
\item{} 
{\it The demonstration of ubiquitous substructure in the 
outer Milky Way,} present in both kinematics and chemical
compositions, probable signatures of hierarchical buildup of
the stellar halo from smaller components.
\item{} 
{\it Demonstration of the common origin of dynamical 
asteroid families,} with distinctive colors indicating similar
composition and space weathering.
\item{} 
{\it Precision measurement of the luminosity distribution of
quasars,} mapping the rise and fall of quasars and the growth of
the supermassive black holes that power them.
\item{} 
{\it Precision measurements of large scale galaxy clustering,}
leading to powerful constraints on the matter and energy contents of 
the Universe and on the nature and origin of the primordial fluctuations
that seeded the growth of cosmic structure.
\item{} 
{\it Precision measurement of early structure with the 
Lyman-$\alpha$ forest,} yielding precise constraints on the
clustering of the underlying dark matter distribution
$1.5-3$ Gyr after the big bang.
\item{} 
{\it Detailed characterization of small and intermediate scale
clustering of galaxies} for classes 
defined by luminosity, color, and morphology, 
allowing strong tests of galaxy formation theories
and statistical determination of the relation between galaxies and dark 
matter halos. 
\item{} 
{\it Discovery of many new companions of the Milky Way and Andromeda,}
exceeding the number found in the previous 70 years, and providing
critical new data for understanding galaxy formation in low mass halos.
\item{} 
{\it Discovery of stars escaping the Galaxy,} ejected by
gravitational interactions with the central black hole, providing
information on the conditions at
the Galactic Center and on the shape, mass, and total extent of
the Galaxy's dark matter halo. 
\item{} 
{\it Discovery of acoustic oscillation signatures in the clustering of 
galaxies,} the first
clear detection of a long-predicted cosmological signal,
opening the door to a new method of cosmological measurement that
is the key to the BOSS survey of SDSS-III.
\item{} 
{\it Measurements of the clustering of quasars over a wide range
of cosmic time,} providing critical constraints on the dark matter
halos that host active black holes of different luminosities at
different epochs.

\end{itemize}

Half of these achievements were among the original ``design goals'' of the SDSS,
but the other half were either entirely unanticipated or not expected to be
nearly as exciting or powerful as they turned out to be. The SDSS and SDSS-II
have enabled systematic investigation and ``discovery'' science in nearly equal
measure, and we expect that tradition to continue with SDSS-III.


Funding for SDSS-III has been provided by the Alfred P. Sloan Foundation, the
Participating Institutions, the National Science Foundation, and the U.S.
Department of Energy. The SDSS-III web site is http://www.sdss3.org/.

SDSS-III is managed by the Astrophysical Research Consortium for the
Participating Institutions. The SDSS-III Collaboration is still growing; at
present, the Participating Institutions are the University of Arizona, the
Brazilian Participation Group, University of Cambridge, University of Florida,
the French Participation Group, the German Participation Group, the Joint
Institute for Nuclear Astrophysics (JINA), Johns Hopkins University, Lawrence
Berkeley National Laboratory, Max Planck Institute for Astrophysics (MPA), New
Mexico State University, New York University, Ohio State University University
of Portsmouth, Princeton University, University of Tokyo, University of Utah,
Vanderbilt University, University of Virginia, and the University of Washington.

\end{document}